\documentclass[a4paper,11pt]{article}
\usepackage{pos}
\usepackage{comment}
\usepackage{xfrac}
\def\mcmule{{{\sc McMule}}}
\newcommand{\beq}{\begin{equation*}}
\newcommand{\eeq}{\end{equation*}}
\newcommand{\bea}{\begin{eqnarray*}}
\newcommand{\eea}{\end{eqnarray*}}

\newcommand{\bc}{\begin{cases}}
\newcommand{\ec}{\end{cases}}
\newcommand{\be}{\begin{enumerate}[a)]}
\newcommand{\ee}{\end{enumerate}}
\newcommand{\bmat}{\begin{bmatrix}}
\newcommand{\emat}{\end{bmatrix}}

\title{NNLO QED corrections and inelastic two-photon exchange in elastic lepton–proton scattering with McMule}
\ShortTitle{NNLO QED corrections and inelastic two-photon exchange}

\author*[a]{Matteo Ronchi}

\affiliation[a]{Institut f\"ur Kernphysik and PRISMA\textsuperscript{++} Cluster of Excellence,\\ Johannes Gutenberg-Universit\"at, Mainz, D-55099, Germany}

\emailAdd{ronchima@uni-mainz.de}

\abstract{We supplement \mcmule,~a Monte Carlo framework that provides QED corrections up to next-to-next-to-leading order (NNLO) in the fine structure constant $\alpha$, with the inelastic two-photon exchange (TPE) contribution, based on the data-driven dispersive calculation of Tomalak, Pasquini and Vanderhaeghen \cite{Tomalak:2017shs}, which we extend to the $Q^2\rightarrow 0$ limit that enables applications in the low-$Q^2$ region relevant for upcoming low-energy experiments. We present results for two experimental settings: the future MAGIX@MESA experiment and the ISR@MAMI measurement. For MAGIX@MESA kinematics, the inelastic TPE is small, comparable to the proton-side NLO corrections and subdominant to both the leptonic radiative corrections and the full NNLO QED corrections once a 1 MeV photon-energy cut is applied. For the ISR@MAMI kinematics, NNLO QED corrections modify the radiative tail below the elastic peak at the 10\% level.}

\FullConference{International Workshop on Low Energy Electron Positron Physics at Jefferson Lab (LEEPP2026)\\
23--27 March 2026\\
Thomas Jefferson National Accelerator Facility, Virginia, USA\\}

\begin{document}
\maketitle
\section{Introduction}
\begin{figure}[htbp]
\centering
 \includegraphics[width=0.39\textwidth]{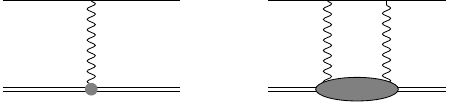}
	\caption{One-photon-exchange ($\mathcal{M}_{1\gamma}$) and two-photon-exchange ($\mathcal{M}_{2\gamma}$) diagrams.}
	\label{fig:OPETPE}
\end{figure}

Scattering experiments probe the structure of the target. In particular, from a measurement of elastic lepton-proton scattering cross sections, we can determine the proton electric and magnetic Sachs form factors (FFs), $G_E(Q^2)$ and $G_M(Q^2)$, functions of the space-like momentum transfer $t=-Q^2$.   
In the non-relativistic limit and in the Breit frame, $G_E$ and $G_M$ are the Fourier transforms of the proton's charge and magnetization densities, respectively.
The FFs describe the electromagnetic interaction of a single photon with the proton (left side of Fig.~\ref{fig:OPETPE}); therefore, in order to extract them, the experimentally measured cross section has to be stripped of interactions with two and more photons --- radiative corrections (Refs.~\cite{Maximon:2000hm, Mo:1968cg, Ent:2001hm, Gramolin:2014pva, Gerasimov:2015aoa, Bystritskiy:2006ju, Bardin:1976qa, Blumlein:2012bf}). This is particularly relevant for upcoming high-precision experiments such as MAGIX (MAinz Gas Injection Target eXperiment) at the future Mainz Energy-recovering Superconducting Accelerator (MESA) facility in Mainz, which plans to perform determinations of both $G_E$ and $G_M$ at low momentum transfer reaching $Q^2 \approx 10^{-5}\,$ GeV$^2$ \cite{Schlimme:2024eky}.
Among the radiative corrections beyond leading order (LO), the two-photon exchange (TPE) contribution (right side of Fig.~\ref{fig:OPETPE}) is particularly important, as it often carries a large relative uncertainty. This contribution has a long history of study: in 2003 Guichon and Vanderhaeghen \cite{Guichon:2003qm} noticed that the TPE can partially solve the discrepancy between the proton FF ratio $\sfrac{G_E}{G_M}$ extracted via the polarization transfer method \cite{Akhiezer:1968ek, Akhiezer:1973xbf} and via the Rosenbluth separation \cite{Rosenbluth:1950yq}, as observed in experiments \cite{JeffersonLabHallA:1999epl, Gayou:2001qt,Punjabi:2005wq,Puckett:2011xg}.
Subsequent theoretical works \cite{Blunden:2003sp,Arrington:2011dn} and dedicated experiments (VEPP-3 \cite{Rachek:2014fam}, CLAS \cite{Adikaram:2014ykv}, OLYMPUS \cite{Henderson:2016dea}) confirmed this picture up to $Q^2\approx 2$ GeV$^2$, via the ratio of $e^+ p \to e^- p$ elastic cross sections.
In this article, we focus on radiative corrections to $lp$ scattering relevant to both past and future experiments, such as the A1@MAMI and the upcoming MAGIX@MESA. In Section \ref{RCandMule}, we briefly review \mcmule~(Monte Carlo for MUons and other LEptons) \cite{Banerjee:2020rww}, a framework for fully differential higher-order QED corrections, focusing on the radiative corrections for $lp\rightarrow lp$ contained within it. We highlight its benefits in this context, such as the computation of QED radiative corrections up to next-to-next-to-leading order (NNLO) in the fine structure constant $\alpha$~\cite{Engel:2023arz}.
In  Section \ref{inTPE}, we discuss how we supplement \mcmule~with the inelastic TPE --- which differs from the elastic contribution in the intermediate hadronic state.
Finally, in Section \ref{results}, we show results of our calculations for the kinematics of the future experiment MAGIX@MESA in Section \ref{magix}, and for the Initial State Radiation (ISR) experiment at MAMI \cite{Mihovilovic:2019jiz} in Section \ref{isr}. We study the impact of the TPE (elastic and inelastic) relative to NNLO QED corrections.
We conclude with a short summary in Section~\ref{conclusions}.
\section{McMule: radiative corrections up to NNLO}
\label{RCandMule}
Early calculations of QED radiative corrections to $lp$ scattering \cite{Mo:1968cg,Maximon:2000hm,Vanderhaeghen:2000ws} were carried out up to an NLO accuracy, neglecting the lepton mass, with TPE  absent or partially included. Data from experiments, such as the A1 experiment at MAMI \cite{A1:2010nsl}, have been analyzed using these radiative corrections.
\mcmule~\cite{Banerjee:2020rww} is a Monte Carlo framework developed to advance the low-energy frontier of processes involving leptons by providing QED corrections up to NNLO. It takes as input 
the relevant matrix elements together with the process kinematics and experimental 
cuts, performs a fully differential phase-space integration using 
$\mathrm{FKS}^\ell$ subtraction \cite{Engel:2019nfw} to handle infrared 
singularities, and returns decay rates or differential cross sections as a function of any suitable
chosen quantity, such as the scattering angle.

\begin{figure}[htbp]
\centering
 \includegraphics[width=0.71\textwidth]{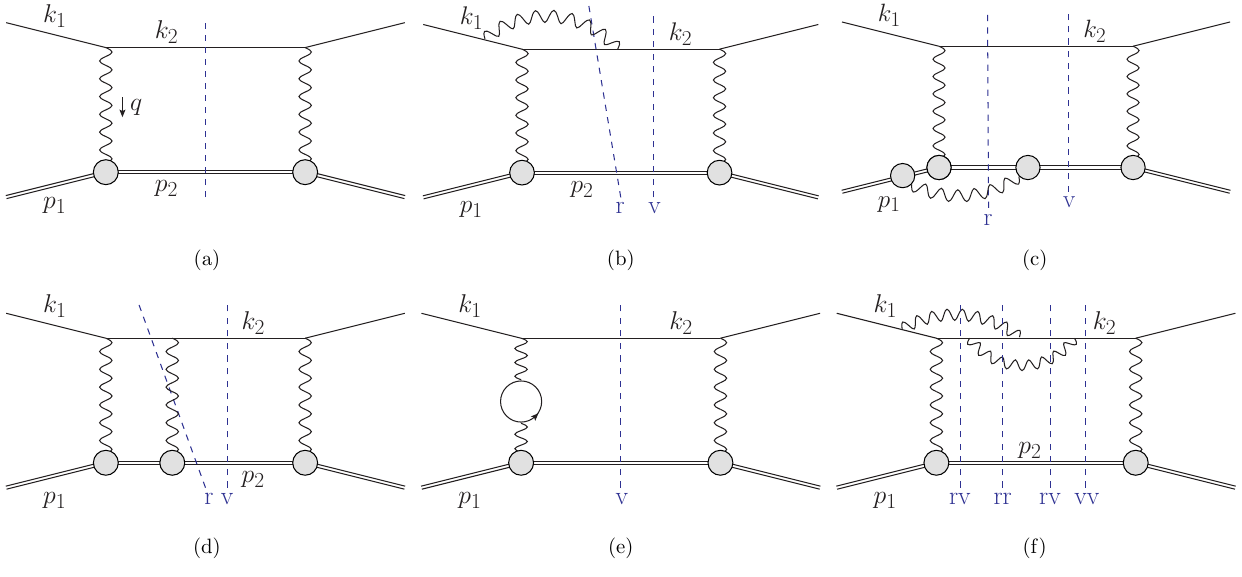}
	\caption{Schematic illustration of representative contributions to the squared amplitude for $lp$ scattering. The leptons (protons) are depicted as single (double) lines and the grey blobs represent the FFs. (a) the squared LO amplitude, (b) and (c) the contributions to the squared NLO amplitude with photon radiation restricted to the lepton or proton line, (d) a contribution to the squared NLO amplitude with elastic two-photon exchange (TPE), (e) NLO correction from vacuum polarization (VP) and (f) an example contribution to the squared NNLO amplitude.
The dashed vertical lines indicate the cut through the diagram, separating it into two conjugate sub-amplitudes. A virtual (v) cut places all photons on one side, representing the interference of a loop correction with the lower-order amplitude. A real (r) cut passes through an emitted photon line, representing the squared modulus of a real-photon-emission amplitude. In the last panel, which involves two photon lines, the cuts are labelled "vv" (the double-virtual interference), "rr" (the doubly-real emission squared), and "rv" (the real–virtual interference and its conjugate). Figures taken from \cite{Afanasev:2023gev}.}
	\label{fig:FeynDiag}
\end{figure}

For $lp$ scattering specifically, \mcmule~provides complete NNLO QED corrections, including the TPE, and does not set the lepton mass to zero. The only exception is the inelastic TPE, which is currently incorporated via external tabulated input rather than computed natively. So far, only its virtual part, i.e. where no photons are emitted, has been considered (see the discussion in Section~\ref{inTPE}).
The complete implementation and discussion of QED radiative corrections are presented in \cite{Engel:2023arz}. A schematic illustration of the contributions of the squared amplitudes is shown in Fig.~\ref{fig:FeynDiag}.

Concerning the TPE, the current implementation in \mcmule~includes only the elastic contribution, where the intermediate hadronic state is a proton, 
modeled with a standard dipole form for $G_E$ and $G_M$. 
The inelastic TPE, which 
accounts for intermediate states beyond the nucleon, is the subject of 
Section~\ref{inTPE}. Since the inelastic TPE contribution 
as discussed here has been calculated only for $e^- p$ scattering \cite{Tomalak:2017shs}, the phenomenological results presented below are restricted to the electron case. The dominant lepton line corrections at NNLO are included without approximation (i.e., with the full electromagnetic proton form factors) when only the leptonic line is involved [e.g., Fig.~\ref{fig:FeynDiag} (f)]; in all other corrections, where photons couple to both the lepton and proton lines, the proton is treated as pointlike \cite{Engel:2023arz}.
\section{Inelastic TPE}
\label{inTPE}
In this section, we discuss the input of inelastic TPE from the work of 
Tomalak, Pasquini and Vanderhaeghen \cite{Tomalak:2017shs}, including its applicability and how it is supplemented inside \mcmule.~
In regard to TPE, the leading correction to the cross section is
\begin{equation}
    \delta_{2\gamma}=\frac{2 \,{\rm{Re}} \left( \mathcal{M}_{1\gamma}^\dagger \mathcal{M}_{2\gamma}\right)}{|\mathcal{M}_{1\gamma}|^2},
\end{equation}
following from the interference between the Born amplitude $\mathcal{M}_{1\gamma}$ and the TPE amplitude $\mathcal{M}_{2\gamma}$. 
For $e^-p$ scattering in the massless lepton limit, the process is fully described by three generalized amplitudes $\mathcal{G}_1$, $\mathcal{G}_2$, and 
$\mathcal{F}_3$~\cite{Tomalak:2016vbf}. These amplitudes can be combined into a 
TPE correction to the unpolarized cross section, defined as
\begin{align}
\label{delta2gepep}
\delta_{2\gamma}&=\frac{2}{G_M^2+\frac{\varepsilon}{\tau}G_E^2} \Big( G_M {\rm{Re}}\mathcal{G}_1^{2\gamma}+\frac{\varepsilon}{\tau}G_E {\rm{Re}}\mathcal{G}_2^{2\gamma} +G_M(\varepsilon-1)  \frac{\nu}{M_p^2}{\rm{Re}}\mathcal{F}_3^{2\gamma}  \Big),
\end{align}
where $\tau$ is the dimensionless momentum transfer, $\nu$ is the crossing-symmetric variable  and $\varepsilon$ is the photon polarization parameter (it varies between $0$ for backward scattering and $1$ for forward scattering):
\begin{align}
\label{relations1}
\tau=\frac{Q^2}{4M_p^2}; \,\, \nu=\frac{s-u}{4}; \,\,\varepsilon=\frac{\nu^2-M_p^4 \tau (1+\tau)}{\nu^2+M_p^4 \tau (1+\tau)}.
\end{align}
The authors of Ref.~\cite{Tomalak:2017shs} evaluated $\delta_{2\gamma}$ within a dispersive framework, accounting for the $\pi N$ intermediate state contributions using the phenomenological
MAID fit \cite{Drechsel:1998hk,Drechsel:2007if} as an input, obtaining amplitudes for the momentum transfer region from $0.005$ GeV$^2$ to $1$ GeV$^2$.

The TPE prediction of Ref.~\cite{Tomalak:2017shs} is available as an interpolation table (private communication) as a function of $Q^2$ and $\varepsilon$. However, the table is available only from $Q^2=0.005$ GeV$^2$, whereas, as explained above, experiments such as MAGIX probe significantly lower momentum transfer. To extend the table to $Q^2 = 0$, we use the following argument from Tomalak's PhD thesis~\cite{Tomalak:2016kyd}: the amplitudes $\mathcal{G}_1^{2\gamma}$ and $\mathcal{G}_2^{2\gamma}$ vanish at $Q^2 = 0$, see Eqs.~(F.10)-(F.11) of \cite{Tomalak:2016kyd}, while for the amplitude $\mathcal{F}_3^{2\gamma}$, the leading behavior near $Q^2 = 0$ is $a\log Q + b$ \cite{Tomalak:2016kyd}. However, the $\mathcal{F}_3^{2\gamma}$ contribution to the unpolarized cross section can be set to 0 at $Q^2=0$ due to the multiplicative kinematic prefactor $(\varepsilon - 1)$ appearing in Eq.~\eqref{delta2gepep}. In fact, at MAGIX the beam energy (and therefore $s$) is fixed. Moreover, once $Q^2$ is fixed, $\nu$ is determined as well. From Eq.~\eqref{relations1}, at $Q^2=0$, $\varepsilon$ becomes $\nu/\nu$, and since $\nu$ is fixed, this ratio equals $1$. As a consequence, at $Q^2=0$, $(\varepsilon-1)\to 0$.
We conclude that the inelastic contribution $\delta_{2\gamma}$ can be set to zero for all $\varepsilon$ at $Q^2=0$. Please note that this argument holds only at the level of $\delta_{2\gamma}$, where the explicit $(\varepsilon-1)$ factor appears.
To supplement this contribution inside \mcmule, we need to have tables of $\delta_{2\gamma}$ as a function of the momentum transfer $Q^2$ for a given beam energy.

\begin{figure}[htbp]
\centering
 \includegraphics[width=0.65\textwidth]{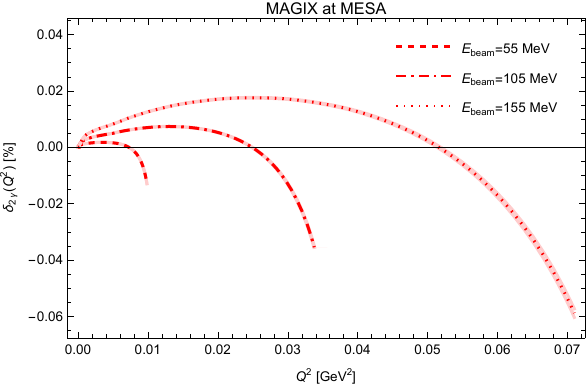}
	\caption{Inelastic TPE correction $\delta_{2\gamma}$ (from Ref.~\cite{Tomalak:2017shs}) as a function of $Q^2$ 
for kinematic conditions relevant to the MAGIX@MESA experiment with the corresponding error band.}
	\label{fig:delta2gmagix}
\end{figure}

In \mcmule, the experimental setup, including the lepton beam energy $E_b$, is specified in a user file~\cite{Ulrich2022-ug}. Inside the user file, through a piecewise bilinear (linear-spline) interpolation in $Q^2$ and $\varepsilon$, we construct from these tables a continuous function $\delta_{2\gamma}(Q^2,\varepsilon)$, together with its upper and lower uncertainty bands. For a fixed $E_b$, we observe that $\varepsilon$ is uniquely determined by $Q^2$ alone, in fact,
\begin{equation}
    \nu(s,Q^2)=\frac{2s-2M_p^2-Q^2}{4}; \,\, s(E_b)=M_p^2+2M_pE_b.
\end{equation}
Evaluating $\delta_{2\gamma}(Q^2,\varepsilon)$ along this kinematic curve [$\varepsilon(Q^2)$] yields $\delta_{2\gamma}(Q^2)$ for the beam energy specified in the user file, which \mcmule\ uses to compute the inelastic TPE contribution to the cross section. Figure~\ref{fig:delta2gmagix} shows the resulting $\delta_{2\gamma}$ as a function of $Q^2$ for three kinematics relevant to the future MAGIX@MESA experiment, with $E_{b}=55$, $105$, and $155$ MeV together with their uncertainty band, showing that for these kinematics $\delta_{2\gamma}$ is a small contribution (around 0.01 \%). 
\section{Results}
\label{results}
This section presents results for $e^-p$ scattering up to NNLO QED, for two experimental settings: the future MAGIX@MESA (Section~\ref{magix}) and the recent ISR@MAMI (Section~\ref{isr}), similar to what has been done for the upcoming MUSE experiment for $\mu p$ scattering in Ref.~\cite{Engel:2023arz}.
\subsection{Predictions for MAGIX@MESA}
\label{magix}
We choose a beam energy of $E_b = 105$~MeV. In order to select 
elastic scattering events and suppress radiative contributions, the experimental 
physical cuts applied are: a minimum momentum of the outgoing electron $p_\text{min} = 73.5$~MeV, scattering angle $\theta \in [15^\circ, 165^\circ]$ and photon-energy cut $E_\gamma = 1$~MeV, meaning that all events in which a photon with energy larger than 1~MeV is produced are discarded.

\begin{figure}[h]
\centering
 \includegraphics[width=0.75\textwidth]{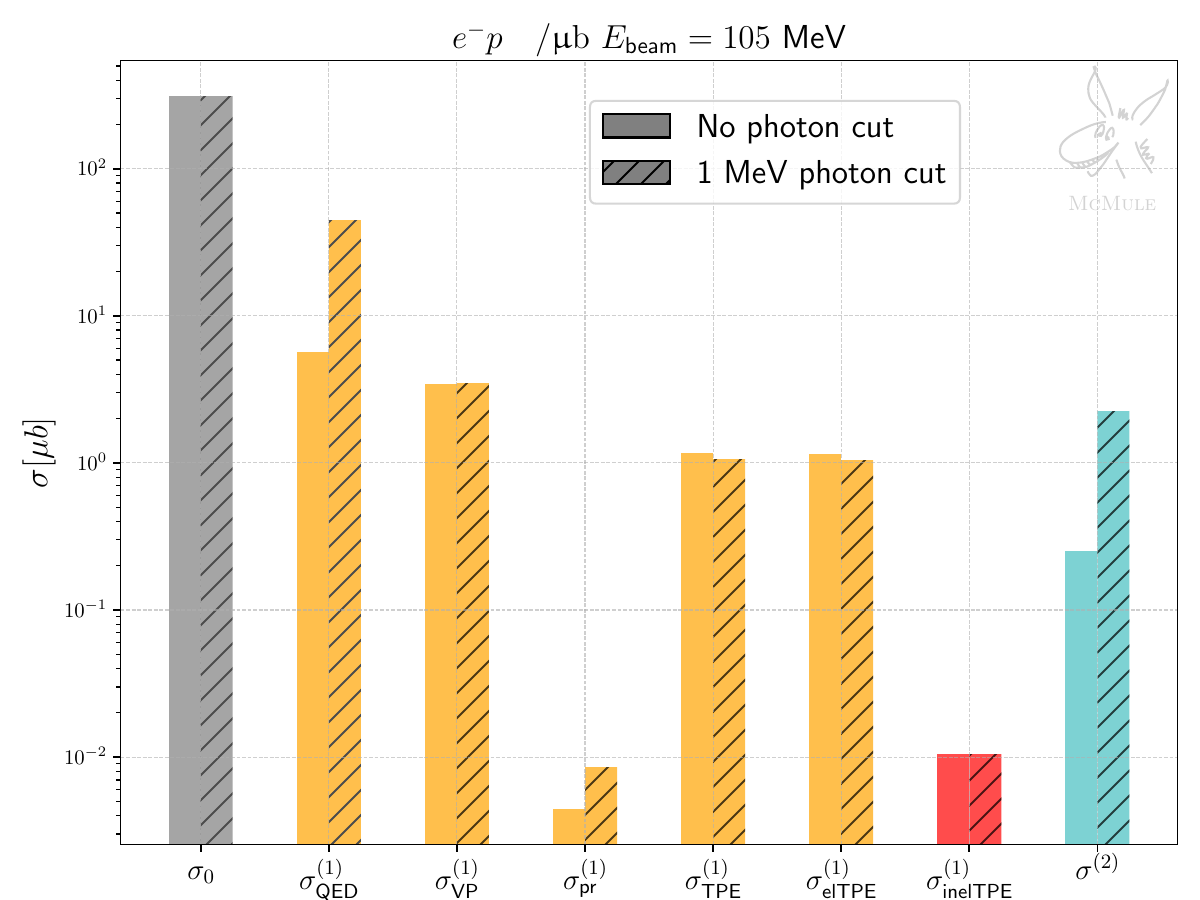}
	\caption{Contributions to the elastic $e^-p$ total cross section at 
$E_b = 105$~MeV with MAGIX@MESA cuts. Each pair of columns 
shows a given contribution without (solid) and with (hatched) the 
$E_\gamma = 1$~MeV photon-energy cut. From left to right: the LO term $\sigma_0$ (Fig.~\ref{fig:FeynDiag}a), the NLO leptonic correction $\sigma^{(1)}_\text{QED}$ (Fig.~\ref{fig:FeynDiag}b), vacuum polarization $\sigma^{(1)}_\text{VP}$ (Fig.~\ref{fig:FeynDiag}e), the proton-side vertex correction $\sigma^{(1)}_\text{pr}$ (Fig.~\ref{fig:FeynDiag}c), the total two-photon-exchange contribution ($\sigma^{(1)}_\text{TPE}$), its elastic component $\sigma^{(1)}_\text{elTPE}$ (Fig.~\ref{fig:FeynDiag}d), its inelastic component discussed in Section~\ref{inTPE} ($\sigma^{(1)}_\text{inelTPE}$), and the full NNLO QED corrections ($\sigma^{(2)}$).}
	\label{fig:histcutnoncut}
\end{figure}

Figure~\ref{fig:histcutnoncut} shows the impact of the photon-energy cut on the different contributions to elastic $e^-p$ scattering, grouped in pairs of bars: the solid bar indicates the result without any photon-energy cut, while the hatched bar corresponds to the result obtained with an $E_\gamma = 1$~MeV cut.
From this comparison we draw two observations. The inelastic TPE is small 
in these kinematics: its contribution is of the same order of magnitude as 
the proton-side NLO corrections, and well below the dominant leptonic 
radiative corrections. Moreover, the NNLO QED corrections become larger 
than the total TPE contribution when the cut is applied. We observe that the inelastic TPE contribution is unchanged by the photon-energy cut, as expected since only its virtual part, with no real-photon emission, is considered in this work.
\subsection{NNLO effects in the ISR@MAMI radiative tail}
\label{isr}
\begin{figure}[htbp]
\centering
 \includegraphics[width=0.74\textwidth]{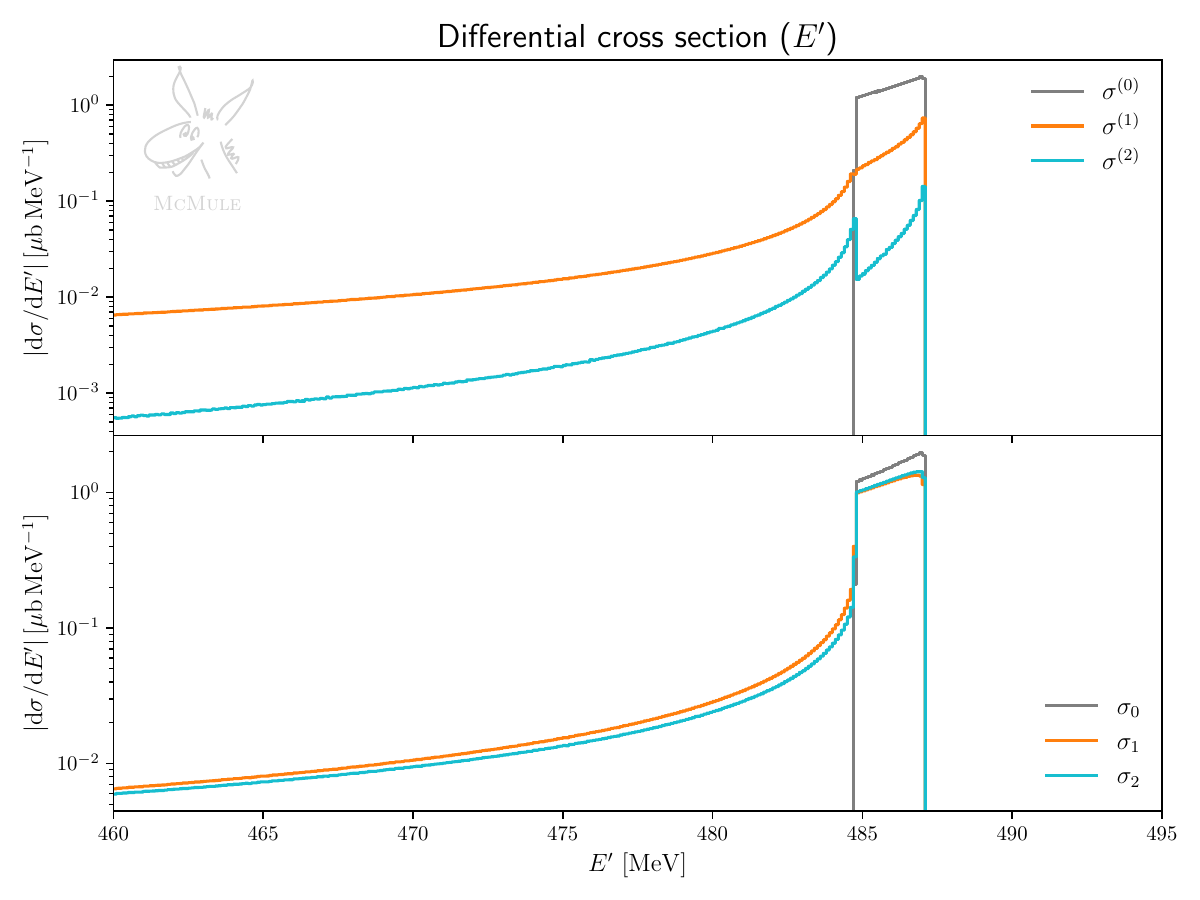}
	\caption{Differential cross section as a function of the scattered 
electron energy $E'$ for ISR@MAMI kinematics at 
$E_b = 495$~MeV. Upper panel: individual LO, NLO, and NNLO 
contributions $\sigma^{(i)}$. Lower panel: cumulative cross section 
$\sigma_k = \sum_{i=0}^{k}\sigma^{(i)}$ up to order $k$. The NNLO 
corrections affect the radiative tail before the elastic peak at the 
$10 \%$ level.}
	\label{fig:radiativetail}
\end{figure}

In this subsection, we show the impact of NNLO QED corrections in the 
ISR experiment conducted at the MAMI accelerator in Mainz \cite{Mihovilovic:2019jiz}. 
The beam energy is $E_b = 495$~MeV, and the experimental cuts 
applied are a minimum momentum of the outgoing electron $p_\text{min} = 280$~MeV 
and a scattering angle $\theta \in [15^\circ \pm 15~\text{mrad}]$.
Figure~\ref{fig:radiativetail} shows two panels. The upper panel displays the absolute value of the differential cross section as a function of the scattered electron energy ($E'$) for the LO, NLO and NNLO contributions ($\sigma^{(i)}$). The lower panel shows the cumulative cross section up to a given order, defined as ($\sigma_{k}=\sum_{i=0}^{k}\sigma^{(i)}$). The NNLO QED corrections have a significant impact on the radiative tail, i.e. the region of $E'$ below the elastic peak, where they modify the cross section at the level of 10\%.
\section{Conclusions}
\label{conclusions}
We have incorporated the inelastic TPE correction of Tomalak, Pasquini and Vanderhaeghen \cite{Tomalak:2017shs}, based on the dispersive treatment of the $\pi N$ intermediate state, into the \mcmule~framework via a tabulated interpolation in $Q^2$ and $\varepsilon$. This allows a direct comparison of the inelastic TPE against the full set of NNLO QED corrections available in \mcmule. For MAGIX@MESA kinematics, the inelastic TPE contribution is small, comparable in size to the purely protonic NLO corrections and well below the dominant leptonic radiative corrections, while NNLO QED corrections become the dominant effect once a photon-energy cut is applied. For ISR@MAMI kinematics, NNLO corrections significantly shape the radiative tail at the 10\% level. Together, these results highlight the importance of a complete, fully differential treatment of radiative corrections for upcoming high-precision electron-proton scattering experiments.

\section*{Acknowledgments}
\label{acknowledgments}
This work was supported by the Deutsche Forschungsgemeinschaft (DFG, German Research Foundation) through the Collaborative Research Center ``CRC1660:
Hadrons and Nuclei as discovery tools'' (Project ID
514321794).
A special thanks goes to my group, who actively contributed to the realization of this work. In particular, I thank my supervisor F. Hagelstein for her support and guidance, as well as for the scientific discussion and V. Lensky for reviewing the manuscript and for his valuable advice.
We thank O.~Tomalak and M.~Vanderhaeghen for providing their results from \cite{Tomalak:2017shs} and useful discussions, M.~Vanderhaeghen for his suggestion to study the effects of the NNLO corrections to the radiative tail.
We thank S.~Schlimme, H.~Merkel and M.~Mihovilovič for valuable discussions regarding A1 and MAGIX. We thank all the \mcmule~team --- in particular M.~Rocco, S.~Kollatzsch, A.~Signer and Y.~Ulrich for their helpful support.


\begin{thebibliography}{99}

\bibitem{Tomalak:2017shs}
O.~Tomalak, B.~Pasquini and M.~Vanderhaeghen,
\textit{Two-photon exchange contribution to elastic $e^-$-proton scattering: Full dispersive treatment of $\pi N$ states and comparison with data},
Phys.\ Rev.\ D \textbf{96} (2017) 096001
[1708.03303].

\bibitem{Maximon:2000hm}
L.~C.~Maximon and J.~A.~Tjon,
\textit{Radiative corrections to electron proton scattering},
Phys.\ Rev.\ C \textbf{62} (2000) 054320
[0002058].

\bibitem{Mo:1968cg}
L.~W.~Mo and Y.~S.~Tsai,
\textit{Radiative Corrections to Elastic and Inelastic $ep$ and $\mu p$ Scattering},
Rev.\ Mod.\ Phys.\ \textbf{41} (1969) 205.

\bibitem{Ent:2001hm}
R.~Ent, B.~W.~Filippone, N.~C.~R.~Makins, R.~G.~Milner, T.~G.~O'Neill and D.~A.~Wasson,
\textit{Radiative corrections for $(e,e'p)$ reactions at GeV energies},
Phys.\ Rev.\ C \textbf{64} (2001) 054610.

\bibitem{Gramolin:2014pva}
A.~V.~Gramolin, V.~S.~Fadin, A.~L.~Feldman, R.~E.~Gerasimov, D.~M.~Nikolenko, I.~A.~Rachek and D.~K.~Toporkov,
\textit{A new event generator for the elastic scattering of charged leptons on protons},
J.\ Phys.\ G \textbf{41} (2014) 115001
[1401.2959].

\bibitem{Gerasimov:2015aoa}
R.~E.~Gerasimov and V.~S.~Fadin,
\textit{Analysis of approximations used in calculations of radiative corrections to electron-proton scattering cross section},
Phys.\ Atom.\ Nucl.\ \textbf{78} (2015) 69.

\bibitem{Bystritskiy:2006ju}
Y.~M.~Bystritskiy, E.~A.~Kuraev and E.~Tomasi-Gustafsson,
\textit{Structure function method applied to polarized and unpolarized electron-proton scattering: A solution of the $G_E(p)/G_M(p)$ discrepancy},
Phys.\ Rev.\ C \textbf{75} (2007) 015207
[0603132].

\bibitem{Bardin:1976qa}
D.~Y.~Bardin and N.~M.~Shumeiko,
\textit{An exact calculation of the lowest order electromagnetic correction to the elastic scattering},
Nucl.\ Phys.\ B \textbf{127} (1977) 242.

\bibitem{Blumlein:2012bf}
J.~Bl\"umlein,
\textit{The theory of deeply inelastic scattering},
Prog.\ Part.\ Nucl.\ Phys.\ \textbf{69} (2013) 28
[1208.6087].

\bibitem{Schlimme:2024eky}
S.~Schlimme et al.,
\textit{The MESA physics program},
EPJ Web Conf.\ \textbf{303} (2024) 06002
[2402.01027].

\bibitem{Guichon:2003qm}
P.~A.~M.~Guichon and M.~Vanderhaeghen,
\textit{How to reconcile the Rosenbluth and the polarization transfer method in the measurement of the proton form-factors},
Phys.\ Rev.\ Lett.\ \textbf{91} (2003) 142303
[0306007].

\bibitem{Akhiezer:1968ek}
A.~I.~Akhiezer and M.~P.~Rekalo,
\textit{Polarization phenomena in electron scattering by protons in the high energy region},
Sov.\ Phys.\ Dokl.\ \textbf{13} (1968) 572.

\bibitem{Akhiezer:1973xbf}
A.~I.~Akhiezer and M.~P.~Rekalo,
\textit{Polarization effects in the scattering of leptons by hadrons},
Fiz.\ Elem.\ Chast.\ Atom.\ Yadra \textbf{4} (1973) 662.

\bibitem{Rosenbluth:1950yq}
M.~N.~Rosenbluth,
\textit{High energy elastic scattering of electrons on protons},
Phys.\ Rev.\ \textbf{79} (1950) 615.

\bibitem{JeffersonLabHallA:1999epl}
M.~K.~Jones et al. [Jefferson Lab Hall A],
\textit{$G_{Ep}/G_{Mp}$ ratio by polarization transfer in $\vec ep\rightarrow e\vec p$},
Phys.\ Rev.\ Lett.\ \textbf{84} (2000) 1398
[9910005].

\bibitem{Gayou:2001qt}
O.~Gayou, K.~Wijesooriya, A.~Afanasev, M.~Amarian, K.~Aniol, S.~Becher, K.~Benslama, L.~Bimbot, P.~Bosted and E.~Brash, et al.,
\textit{Measurements of the elastic electromagnetic form-factor ratio $\mu_p G_{Ep}/G_{Mp}$ via polarization transfer},
Phys.\ Rev.\ C \textbf{64} (2001) 038202.

\bibitem{Punjabi:2005wq}
V.~Punjabi, C.~F.~Perdrisat, K.~A.~Aniol, F.~T.~Baker, J.~Berthot, P.~Y.~Bertin, W.~Bertozzi, A.~Besson, L.~Bimbot and W.~U.~Boeglin, et al.,
\textit{Proton elastic form-factor ratios to $Q^2$ = 3.5 GeV$^2$ by polarization transfer},
Phys.\ Rev.\ C \textbf{71} (2005) 055202
[Erratum: Phys.\ Rev.\ C \textbf{71} (2005) 069902]
[0501018].

\bibitem{Puckett:2011xg}
A.~J.~R.~Puckett, E.~J.~Brash, O.~Gayou, M.~K.~Jones, L.~Pentchev, C.~F.~Perdrisat, V.~Punjabi, K.~A.~Aniol, T.~Averett and F.~Benmokhtar, et al.,
\textit{Final Analysis of Proton Form Factor Ratio Data at $Q^2 = 4.0$, 4.8 and 5.6 GeV$^2$},
Phys.\ Rev.\ C \textbf{85} (2012) 045203
[1102.5737].

\bibitem{Blunden:2003sp}
P.~G.~Blunden, W.~Melnitchouk and J.~A.~Tjon,
\textit{Two photon exchange and elastic electron proton scattering},
Phys.\ Rev.\ Lett.\ \textbf{91} (2003) 142304
[0306076].

\bibitem{Arrington:2011dn}
J.~Arrington, P.~G.~Blunden and W.~Melnitchouk,
\textit{Review of two-photon exchange in electron scattering},
Prog.\ Part.\ Nucl.\ Phys.\ \textbf{66} (2011) 782
[1105.0951].

\bibitem{Rachek:2014fam}
I.~A.~Rachek et al.,
\textit{Measurement of the two-photon exchange contribution to the elastic $e^{\pm}p$ scattering cross sections at the VEPP-3 storage ring},
Phys.\ Rev.\ Lett.\ \textbf{114} (2015) 062005
[1411.7372].

\bibitem{Adikaram:2014ykv}
D.~Adikaram et al. [CLAS Collaboration],
\textit{Towards a resolution of the proton form factor problem: New electron and positron scattering data},
Phys.\ Rev.\ Lett.\ \textbf{114} (2015) 062003
[1411.6908].

\bibitem{Henderson:2016dea}
B.~S.~Henderson et al. [OLYMPUS Collaboration],
\textit{Hard Two-Photon Contribution to Elastic Lepton-Proton Scattering: Determined by the OLYMPUS Experiment},
Phys.\ Rev.\ Lett.\ \textbf{118} (2017) 092501
[1611.04685].

\bibitem{Banerjee:2020rww}
P.~Banerjee, T.~Engel, A.~Signer and Y.~Ulrich,
\textit{QED at NNLO with McMule},
SciPost Phys.\ \textbf{9} (2020) 027
[2007.01654].

\bibitem{Engel:2023arz}
T.~Engel, F.~Hagelstein, M.~Rocco, V.~Sharkovska, A.~Signer and Y.~Ulrich,
\textit{Impact of NNLO QED corrections on lepton-proton scattering at MUSE},
Eur.\ Phys.\ J.\ A \textbf{59} (2023) 253
[2307.16831].

\bibitem{Mihovilovic:2019jiz}
M.~Mihovilovi{\v{c}} et al.,
\textit{The proton charge radius extracted from the initial-state radiation experiment at MAMI},
Eur.\ Phys.\ J.\ A \textbf{57} (2021) 107
[1905.11182].

\bibitem{A1:2010nsl}
J.~C.~Bernauer et al. [A1 Collaboration],
\textit{High-precision determination of the electric and magnetic form factors of the proton},
Phys.\ Rev.\ Lett.\ \textbf{105} (2010) 242001
[1007.5076].

\bibitem{Vanderhaeghen:2000ws}
M.~Vanderhaeghen, J.~M.~Friedrich, D.~Lhuillier, D.~Marchand, L.~Van Hoorebeke and J.~Van de Wiele,
\textit{QED radiative corrections to virtual Compton scattering},
Phys.\ Rev.\ C \textbf{62} (2000) 025501
[0001100].

\bibitem{Engel:2019nfw}
T.~Engel, A.~Signer and Y.~Ulrich,
\textit{A subtraction scheme for massive QED},
JHEP \textbf{01} (2020) 085
[1909.10244].

\bibitem{Afanasev:2023gev}
A.~Afanasev, J.~C.~Bernauer, P.~Blunden, J.~Bl{\"u}mlein, E.~W.~Cline,
J.~M.~Friedrich, F.~Hagelstein, T.~Husek, M.~Kohl and F.~Myhrer, et al.,
\textit{Radiative corrections: from medium to high energy experiments},
Eur.\ Phys.\ J.\ A \textbf{60} (2024) 91
[2306.14578].

\bibitem{Tomalak:2016vbf}
O.~Tomalak, B.~Pasquini and M.~Vanderhaeghen,
\textit{Two-photon exchange corrections to elastic $e^-$-proton scattering: Full dispersive treatment of $\pi N$ states at low momentum transfers},
Phys.\ Rev.\ D \textbf{95} (2017) no.9, 096001
[1612.07726].

\bibitem{Drechsel:1998hk}
D.~Drechsel, O.~Hanstein, S.~S.~Kamalov and L.~Tiator,
\textit{A unitary isobar model for pion photoproduction and electroproduction on the proton up to 1 GeV},
Nucl.\ Phys.\ A \textbf{645} (1999) 145
[9807001].

\bibitem{Drechsel:2007if}
D.~Drechsel, S.~S.~Kamalov and L.~Tiator,
\textit{Unitary isobar model -- MAID2007},
Eur.\ Phys.\ J.\ A \textbf{34} (2007) 69
[0710.0306].

\bibitem{Tomalak:2016kyd}
O.~Tomalak,
\textit{Two-photon exchange corrections in elastic lepton-proton scattering},
Ph.D.\ Thesis, Johannes Gutenberg-Universit\"at Mainz (2016),
doi:10.25358/openscience-1037.

\bibitem{Ulrich2022-ug}
Y.~Ulrich, P.~Banerjee, A.~Coutinho, T.~Engel, A.~Gurgone, F.~Hagelstein, S.~Kollatzsch, L.~Naterop, M.~Rocco, N.~Schalch, V.~Sharkovska and A.~Signer,
\textit{The McMule Manual},
Zenodo (2022),
doi:10.5281/zenodo.6046769.

\end{thebibliography}
\end{document}